\documentclass[secnumarabic,titlepage,preprint,nofootinbib,floatfix,superscriptaddress]{revtex4-2}
\usepackage{graphicx}
\usepackage{multirow}
\usepackage{amsmath}
\usepackage{slashed}
\usepackage{mathtools}
\usepackage{physics}
\usepackage{dsfont}
\usepackage{subcaption}
\usepackage{soul}
\begin{document}
\title{$D_1(2420)$ and its interactions with a kaon: open charm states with strangeness}

\author{Brenda B. Malabarba}
\email{brenda@if.usp.br}
\affiliation{Universidade de Sao Paulo, Instituto de Fisica, C.P. 05389-970, Sao Paulo,  Brazil.}

\author{K. P. Khemchandani}
 \email{kanchan.khemchandani@unifesp.br}
\affiliation{Universidade Federal de S\~ao Paulo, C.P. 01302-907, S\~ao Paulo, Brazil.}

\author{A.~\surname{Mart\'inez~Torres}}
\email{amartine@if.usp.br}
\affiliation{Universidade de Sao Paulo, Instituto de Fisica, C.P. 05389-970, Sao Paulo,  Brazil.}

\author{E. Oset}
\email{Eulogio.Oset@if.usp.br}
\affiliation{Departamento de F\'isica Te\'orica and IFIC,
Centro Mixto Universidad de Valencia-CSIC,
Institutos de Investigaci\'on de Paterna,
Aptdo. 22085, 46071 Valencia, Spain.}

\preprint{}

\date{\today}
\begin{abstract}
In this work we present an attempt to describe the $X_1(2900)$ found by the LHCb collaboration, in the experimental data on the invariant mass spectrum  of $ D^-K^+$, as a three-meson molecular state of the $K\rho\bar D$ system. We discuss that the interactions in all the subsystems are attractive in nature, with the $\rho \bar D$ interaction generating $\bar D_1(2420)$ and the $K \rho$ resonating as $K_1(1270)$. We find that the system can form a three-body state but with a mass higher than that of $X_1(2900)$. We investigate the $K \rho D$ system too, finding that the three-body dynamics generates an isoscalar state, which can be related to $D_{s1}^*(2860)$, and an exotic isovector state. This latter state has a mass similar to that of the $X_0(2900)$ and $X_1(2900)$ states found by LHCb, but a very small width ($\sim 7.4 \pm 0.9$ MeV) and necessarily requires more than two quarks to describe its properties. We hope that our findings will encourage experimental investigations of the isovector $K \rho D$ state. Finally, in the pursuit of finding a description for $X_1(2900)$, we study the $\pi\bar K^* D^*$ system where $ \bar K^*D^*$ forms $0^+$, $1^+$ and $2^+$ states. We do not find a state which can be associated with $X_1(2900)$.

\end{abstract}

\maketitle

\section{Introduction}
The motivation of the present work is to find a description for $X_1(2900)$, a state with open charm and strange quantum numbers, discovered by LHCb~\cite{LHCb:2020bls,LHCb:2020pxc}. The state was found together with a  spin-parity $0^+$ resonance, $X_0(2900)$,  in the $D^-K^+$ invariant mass.  We shall refer to these states as $X_0$ and $X_1$ in the present manuscript. Such states clearly require more than two quarks to describe their quantum numbers: $C=-1,~S=1$, adding a new task to understanding the nature of the explicitly non $q\bar q$ states found within the last two decades. The masses and widths of the two states are determined in Refs.~\cite{LHCb:2020bls,LHCb:2020pxc} as: $M_{X_0}= 2866\pm 7$ MeV, $\Gamma_{X_0}= 57\pm 13$ MeV, $M_{X_1}= 2904\pm 5$ MeV and $\Gamma_{X_1}= 110\pm 12$ MeV. The isospin of the states is not yet well determined, though different suggestions have been brought forward by different model calculations. 

A variety of descriptions have also been proposed for the structure of $X_i$s, which, naturally, consist of  a compact tetraquark structure, or a molecular nature. Most works agree on attributing an isoscalar $\bar D^* K^*$ quasi bound state to $X_0(2900)$~\cite{Molina:2010tx,Molina:2020hde,Liu:2020nil,Chen:2020aos,Huang:2020ptc,Hu:2020mxp,Agaev:2020nrc,Chen:2021erj,Kong:2021ohg}. An isoscalar compact tetraquark description has also been tested for $X_0$ in Refs.~\cite{Karliner:2020vsi,He:2020jna,Wang:2020xyc,Zhang:2020oze,Wang:2020prk} (as well as for $X_1$ in Refs.~\cite{He:2020jna,Agaev:2021knl}), obtaining the mass but not always the width~\cite{He:2020jna} in good agreement with the data~\cite{LHCb:2020bls,LHCb:2020pxc}.   Further, a compact tetraquark nature has been disfavored in Ref.~\cite{Lu:2020qmp} for both $X_0$ and $X_1$, on the basis of a relativistic quark model. The authors of Ref.~\cite{Albuquerque:2020ugi} suggest that $X_0$ and $X_1$ can be explained as a superposition of a tetraquark and a molecular component. A yet other possibility has been investigated in Ref.~\cite{Liu:2020orv}, indicating that $X_0$ and $X_1$ can arise from a triangular singularity in  $\chi_{c1}K^*\bar D^*$ and $D_{sJ}\bar D_1K^0$  loops, respectively. In the given scenario, the authors of  Refs.~\cite{Burns:2020xne,Chen:2020eyu,Yu:2017pmn} suggest methods and mechanisms to determine the nature of $X_i$'s. 

From the above discussion, it can be noticed that $X_0$ has been studied more than $X_1$.  Since the latter one has spin-parity  $1^-$, it can not be contemplated as a s-wave molecular state of a pair of pseudoscalar/vector mesons. However, one could consider studying a system of an axial and a vector meson interacting in s-wave, which has indeed been done in Refs.~\cite{Dong:2020rgs,He:2020btl,Qi:2021iyv,Chen:2021tad}. All these former works investigate the $\bar D_1 K$ system by writing an effective field for $\bar D_1(2420)$ and relying on the aspects of heavy quark symmetry, but find different results. The interactions between $\bar D_1$ and $K$ have been deduced in Ref.~\cite{Dong:2020rgs} through  vector meson exchange diagrams, which is found to be too weak  to bind the system.  The same formalism is applied to the $\bar D K_1$, $DK_1$ and $D_1 K$ systems, with $K_1$ representing $K_1(1270)$ or $K_1(1400)$, but only $D K_1$ is found to form a bound state. A similar approach is considered in Refs.~\cite{He:2020btl,Qi:2021iyv,Chen:2021tad} but the results obtained are different. It is concluded in Ref.~\cite{He:2020btl}, considering $\bar D K_1$ and $\bar D^*K^*$ as coupled channels, that a large cutoff is required to bind the systems, while the authors of Ref.~\cite{Qi:2021iyv} conclude that, within the uncertainties of their model,  $X_1$ can be interpreted as a $\bar D_1 K$ resonance/bound state. Yet another study is reported in Ref.~\cite{Chen:2021tad},  where $K$-matrices are evaluated with kernels obtained from heavy meson chiral perturbation theory, and an isosinglet, $\bar D_1 K$ molecular interpretation is found to be favorable for $X_1$, discarding a triangular singularity description.

As a summary, it can be said that though a consensus seems to appear on the description of $X_0$, the nature of  $X_1$ is far from clear and further investigations are required. In the present work we investigate a possible explanation for $X_1$, in terms of a three-meson bound  system: $K \rho \bar D$ (or, equivalently, $\bar K\rho  D$).  We solve three-body equations within the static or fixed center approximation (FCA), considering all interactions in $s$-wave and considering $\rho\bar D $ to form a cluster. We first solve Bethe-Salpeter equations for the different subsystems, considering appropriate coupled channels, where $ \rho \bar D$ and coupled channels generate $\bar D_1(2420)$, $K \rho$ (and the respective coupled channels) generates $K_1(1270)$  while $K \bar D$ leads to a weakly attractive amplitude. The $K \rho$ and  $K \bar D$  amplitudes are used as an input to solve the scattering equations for the three-meson system. Our formalism is different to that of Refs.~\cite{Dong:2020rgs,He:2020btl,Qi:2021iyv,Chen:2021tad}, since we have the simultaneous treatment of $\rho \bar D$ as $\bar D_1(2420)$ and $K \rho$ as $K_1(1270)$ in the system. We investigate different total isospins of the system, and find that a state could arise as a consequence of the interactions in the isoscalar configuration of the $K \rho \bar D$ system. However, the mass of such state would be higher than that  determined for $X_1$ in Refs.~\cite{LHCb:2020bls,LHCb:2020pxc}. 

Further, we  investigate the $C=S=+1$ $K \rho D$ system too, where the $KD$ interaction is attractive and forms $D_s(2317)$. In this case, we find an isoscalar state, which can be related with $D_{s1}^*(2860)$, and an  isovector state, which unavoidably requires more than two quarks to describe its properties. The latter exotic state is a  $C=S=+1$ isovector partner of $X_1(2900)$. 

Finally, inspired by the study in Ref.~\cite{Molina:2020hde}, where bound states of  $\bar K^*  D^*$ with spin-parity $0^+$, $1^+$ and $2^+$ have been predicted, we investigate if a pion is added to $\bar K^* D^*$, the resultant three-body system leads to the formation of  bound state(s). It is important to mention that the masses obtained for the $0^+$, $1^+$ and $2^+$ states are $2866$, $2861$ and $2775$~MeV, respectively, in  Ref.~\cite{Molina:2020hde}, and the $0^+$ state has been associated with $X_0$. Our special interest lies in verifying if  the $\bar  K^* D^*$ state with  $J^P=$ $1^+$  of Ref.~\cite{Molina:2020hde} together with a pion forms a vector state, which could be interpreted as the $X_1$ found by LHCb~\cite{LHCb:2020bls,LHCb:2020pxc}. Such a question is motivated by the fact that the masses of $X_0$ and $X_1$ differ by less than the mass of a pion. We do not find a clear state formed around 2900 MeV in the $\pi \bar K^* D^*$ system.

The present manuscript is organized as follows. We first discuss the interactions of $\rho D$ and coupled channels in detail in the next section, showing that $D_1(2420)$ is generated from the underlying dynamics. We show that the properties of $D_1(2420)$ are well described in this model and discuss that meson-meson interactions must give important contributions to  explain the nature of this axial state. In the subsequent section, we discuss the formalism used to study the three-body systems and present the results on $K \rho \bar D$ and $K\rho D$ amplitudes. We dedicate section IV on the discussions of the study of  the $\pi \bar K^*D^*$ system. Finally,  we present a summary and future perspectives of our present work.

\section{Description of $D_1(2420)$ in terms of meson-meson interactions}\label{sec:2}
The  $D_1(2420)$-meson is listed as the lowest mass charmed axial meson in Ref.~\cite{pdg}. Its mass and width are known with a reasonable precision, with the values $2422\pm0.6$ MeV and $31.3\pm 1.9$ MeV, respectively. Interestingly, there exists another axial meson with  a very similar nominal mass, $D_1(2430)$, but with the width of the order of $314\pm29$ MeV.  Keeping this large width in mind, it should be difficult to decide which of the two aforementioned states is the lightest axial with charm. Besides, it is important to recall that in spite of having a very similar mass, same quantum numbers, the two states have very different decay widths. The main decay channel of $D_1(2430)$ is $\pi D^*$ and due to a large phase space available for the decay, the corresponding decay width turns out to be large. Given the similarity in the masses and quantum numbers of $D_1(2420)$ and $D_1(2430)$, but a large difference between their widths, one can infer  that  the two states must have a different nature.

Indeed, on the basis of a coupled channel study of pseudoscalar and vector mesons in Ref.~\cite{gamermann2007}, it was shown that a state, with the properties like those of  $D_1(2420)$, arises from the underlying dynamics. It was also found that the state coupled weakly to the $\pi D^*$ channel, which would explain the small width of  $D_1(2420)$. However, as we discuss below, the work in Ref.~\cite{gamermann2007} needs to be updated such as to better agree with the experimental data.
The formalism in Ref.~\cite{gamermann2007} was built 
using a Lagrangian based on the $SU(4)$ symmetry, broken to $SU(3)$ by suppressing terms in the Lagrangian where the interactions should be driven by the exchange of charmed mesons. Different coupled channels were considered with total charm $+1$ and three poles were found in the complex energy plane, when calculating the modulus squared of the two body $t$-matrix in $I=1/2$. The corresponding pole positions~\cite{gamermann2007} are: $(2311.24 -i 115.68)$~MeV, $(2526.47 - 0.08)$~MeV and $(2750.22 - i 99.91)$~MeV. The authors attempted to identify the pole at $(2526.47 - 0.08)$~MeV with the state $D_1(2420)$ in Ref.~\cite{gamermann2007}, whose mass and width are  $2422.1\pm 0.6$ MeV and  $31.3\pm1.9$ MeV, respectively~\cite{pdg}. Even when the finite widths of the $\rho$ and $K^*$ mesons were considered in  Ref.~\cite{gamermann2007},  and a width of $26$ MeV was obtained for the state identified with $D_1(2420)$,  the mass remained about $100$ MeV above the mass observed experimentally.

In view of the discrepancy between the data and the results presented in Ref.~\cite{gamermann2007}, we update the latter model by (1) considering additional diagrams and (2) fine tuning the parameters of the theory, which are a decay constant  present in the potential and a subtraction constant regularizing the divergent loop function in the Bethe-Salpeter equation. We have verified that both the aforementioned corrections are necessary to well describe the properties of $D_1(2420)$ and any one of the two changes alone is not sufficient.

 In the present work we consider box diagrams  involving the exchange of pseudoscalar mesons (pions), which were not included in Ref.~\cite{gamermann2007}.

\begin{figure}[h!]
    \centering
    \includegraphics[width = 0.4\textwidth]{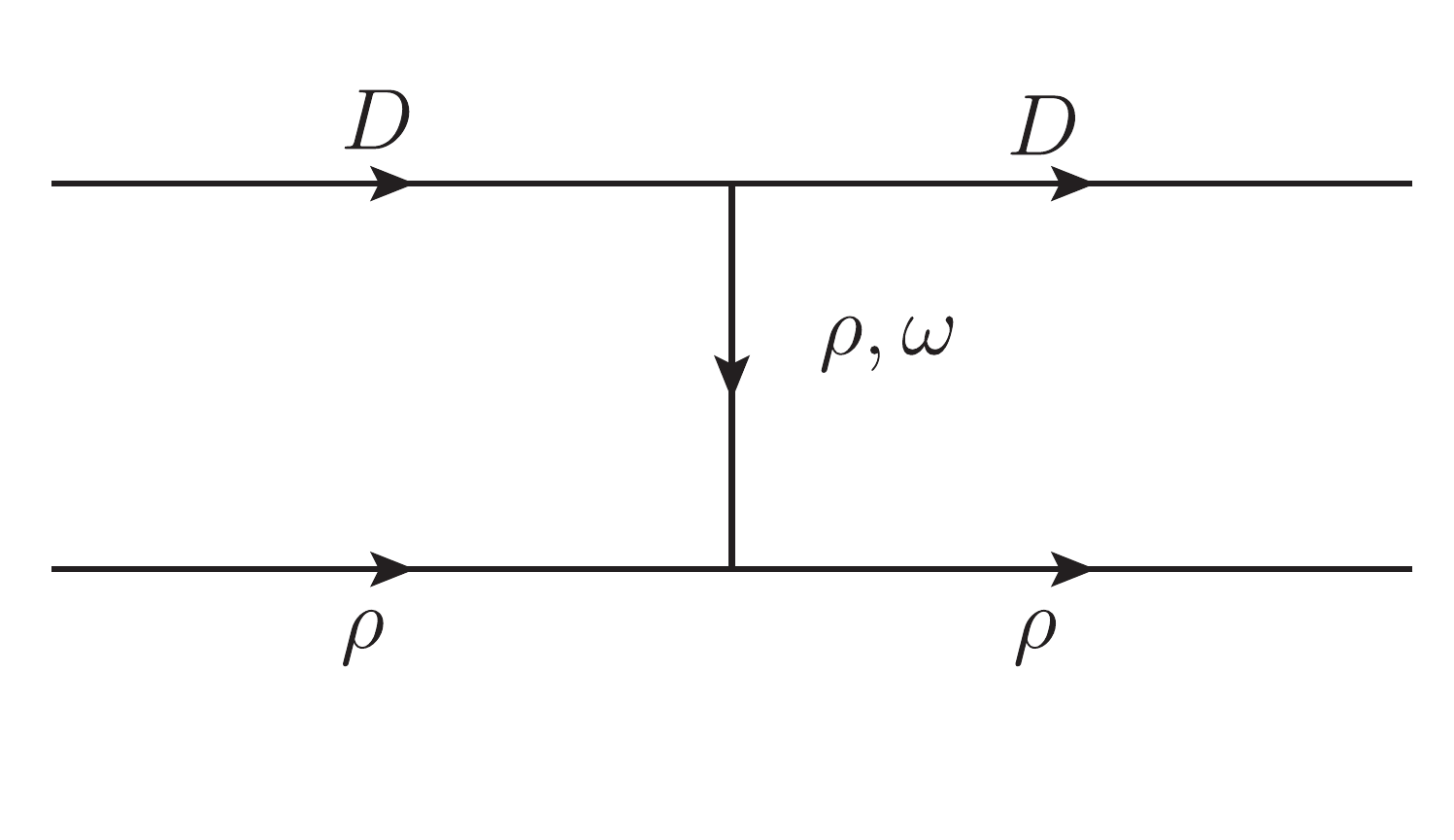}
    \caption{$\rho D$ interaction proceeding through the exchange of vector mesons in the $t$-channel.}
    \label{DrhoHLS}
\end{figure}

\begin{figure}[h!]
    \centering
    \includegraphics[width =  0.5\textwidth]{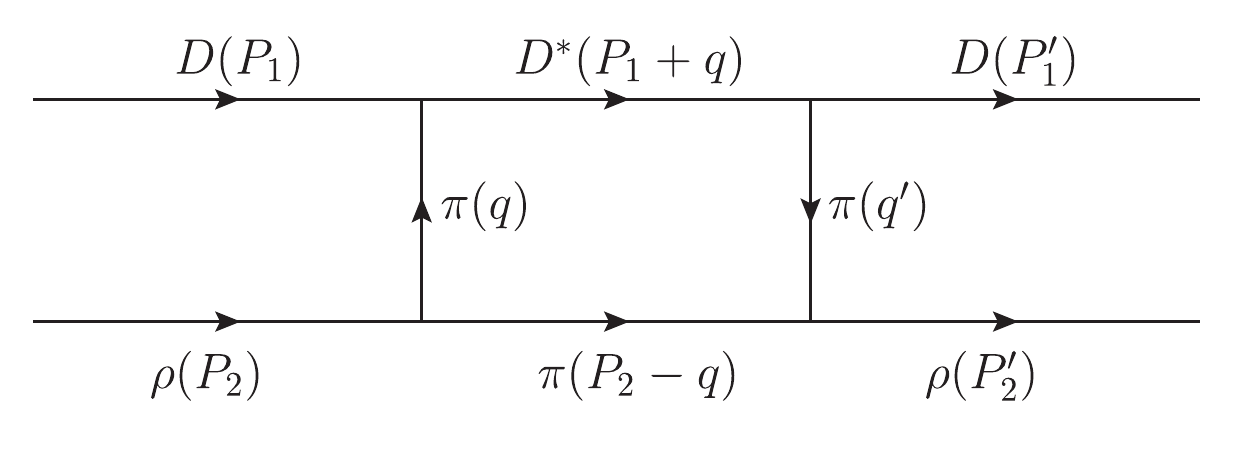}
    \caption{Box diagrams for the $\rho D\to \rho D$ transition in the isospin basis, considering the exchange of pseudoscalar meson, $\pi$, in the $t$-channel.}
    \label{Drhocaixa}
\end{figure}

As shown in Ref.~\cite{Aceti:2014uea}, considering a Lagrangian, like the one used in Ref.~\cite{gamermann2007}, is compatible with considering contributions from diagrams with the exchange of a vector meson in the $t$-channel, as depicted in Fig.~\ref{DrhoHLS},
 with the vertices determined from the Local Hidden Symmetry Lagrangian of Ref.~\cite{Bando:1987br}, and taking the limit $t\to 0$. 
This fact illustrates that  box diagrams, involving exchange of pions (as shown in Fig.~\ref{Drhocaixa}), are an alternative source of contributions to the lowest order pseudoscalar--vector-meson amplitudes, and were not considered in Ref.~\cite{gamermann2007}.

Before proceeding further, it is important to mention that the pole associated with $D_1(2420)$ in Ref.~\cite{gamermann2007} was found to couple strongly to only two channels: $\rho D$ and $D_s\bar K^*$. The couplings for the remaining channels were found to be smaller by about a factor of 5 to 10.  Thus,  we essentially need to consider box diagrams for the $\rho D$ and $D_s\bar K^*$ channels. 

\begin{figure}[h!]
    \centering
    \includegraphics[width = 0.5\textwidth]{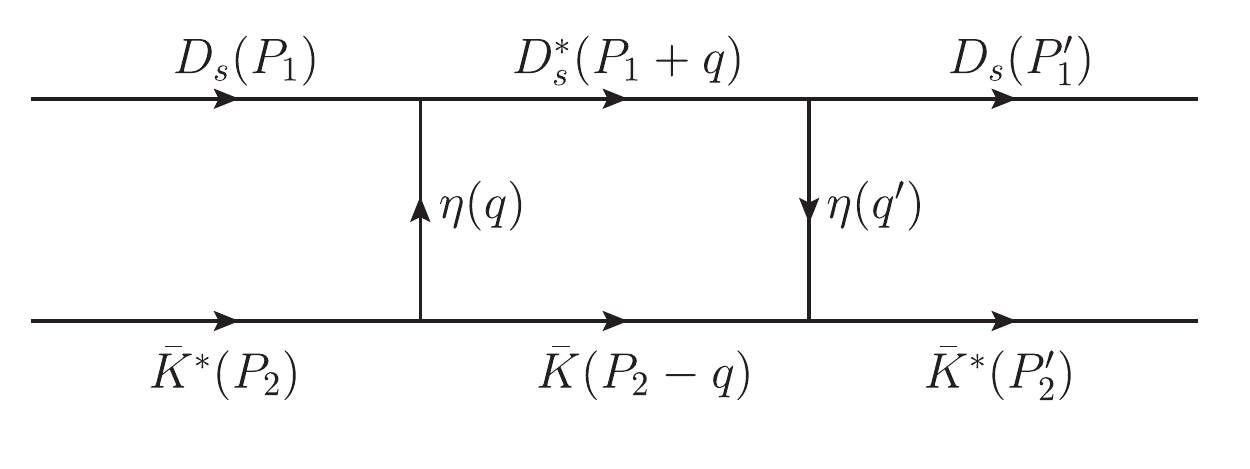}
    \caption{Box diagrams for the transition $D_s\bar K^*\to D_s\bar K^*$ proceeding through the exchange of pseudoscalar mesons in the $t$-channel.}
    \label{DsbarKstrcaixa}
\end{figure}

Furthermore, $D_s\bar K^* \to D_s\bar K^*$ scattering, involving pseudoscalar meson exchange, proceeds through the mechanism depicted in Fig.~\ref{DsbarKstrcaixa}.
As can be seen, the exchanged ($\eta$) particles involved in the loop, contrary to the corresponding ones in Fig.~\ref{Drhocaixa} (pions), are far from being on-shell\footnote{ The threshold $m_{D_s}+m_\eta$ is around 400 MeV far from $m_{D^*_s}$, while $m_\eta+m_{\bar K}$ is about 151 MeV far from $m_{\bar K^*}$.}, and they are heavier when compared to those present in the loop shown in Fig.~\ref{Drhocaixa}. It can be easily verified that the case for the transition between $\rho D$ and $K^* \bar D_s$ is similar. Consequently, we can say that  an important contribution is expected to arise only from the box diagram shown in Fig.~\ref{Drhocaixa}.

To obtain the amplitude for the diagram in Fig.~\ref{Drhocaixa}, we need to determine the contributions from each vertex as well as from the four propagators in the loop.
In order to proceed, we recall that the mesons $D$ and $\rho$ have isospin $1/2$ and $1$, respectively, and, hence,  the  $\rho D$ system can have total isospin $1/2$ or $3/2$.
We are interested in the $\rho D$ system with total isospin $1/2$ (to obtain  a dynamically generated state with the quantum numbers of $D_1(2420)$~\cite{pdg}), which can be written as 
\begin{equation}
    \ket{D \rho, I = 1/2, I_3 = 1/2} = -\sqrt{\frac{2}{3}}\ket{D^0\rho^+} + \sqrt{\frac{1}{3}}\ket{D^+\rho^0},\label{drhostate}
\end{equation}
following the phase convention:
\begin{equation}
    \ket{\rho} = \left(\begin{array}{c}
    -\ket{\rho^+}\\
    \ket{\rho^0}\\
    \ket{\rho^-}
    \end{array}\right),\phantom{naa}\ket{D} = \left(\begin{array}{c}
    \ket{D^+}\\
    -\ket{D^0} 
    \end{array}\right),\phantom{nna}\ket{\bar{D}} = \left(\begin{array}{c}
    \ket{\bar{D}^0}\\
    \ket{D^-} 
    \end{array}\right).\label{convsinal}
\end{equation}

Using Eq.~(\ref{drhostate}), we can obtain the isospin 1/2 amplitude for $D\rho \to D\rho$ as
\begin{align}
   t^{I = 1/2}_{D\rho\to D\rho} 
   = \frac{2}{3}t_{D^0\rho^+\to D^0\rho^+} + \frac{1}{3}t_{D^+\rho^0\to D^+\rho^0} - \frac{2\sqrt{2}}{3}t_{D^0\rho^+\to D^+\rho^0}.\label{nabasedecarga}
\end{align}

As can be seen from Eq. (\ref{nabasedecarga}),  to obtain the  $D\rho\to D\rho$  $t$-matrix in isospin 1/2 we must  determine the amplitudes for the processes: $D^0\rho^+\to D^0\rho^+$, $D^+\rho^0\to D^+\rho^0$, $D^+\rho^0\to D^0\rho^+$.

\begin{figure}[h!]
\centering
\begin{subfigure}{0.45\textwidth}
    \includegraphics[width=\textwidth]{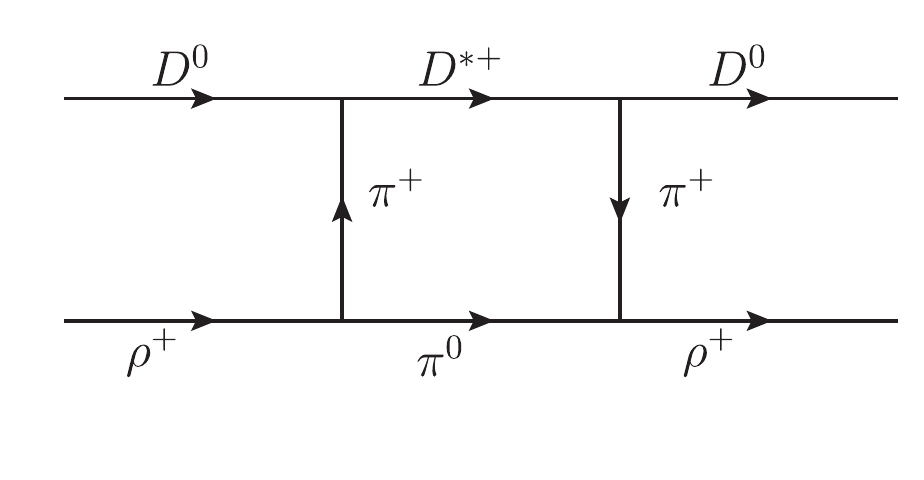}
    \caption{}
    \label{diag11}
\end{subfigure}
\quad
\begin{subfigure}{0.45\textwidth}
    \includegraphics[width=\textwidth]{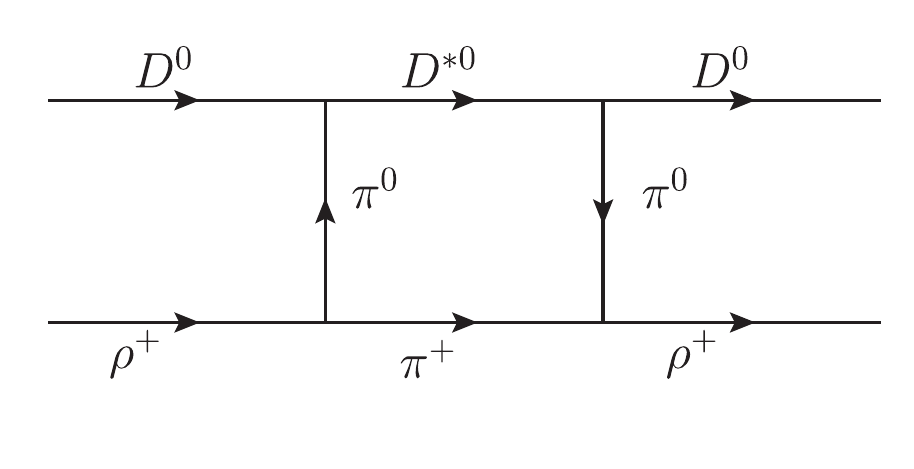}
    \caption{}
    \label{diag12}
\end{subfigure}
\caption{Box diagrams for the transitions $D^0\rho^+\to  D^{*+}\pi^0\to D^0\rho^+$ (left) and $D^0\rho^+\to  D^{*0}\pi^+\to D^0\rho^+$ (right)}
\label{diag1}
\end{figure}

\begin{figure}[h!]
\centering
\begin{subfigure}{0.45\textwidth}
    \includegraphics[width=\textwidth]{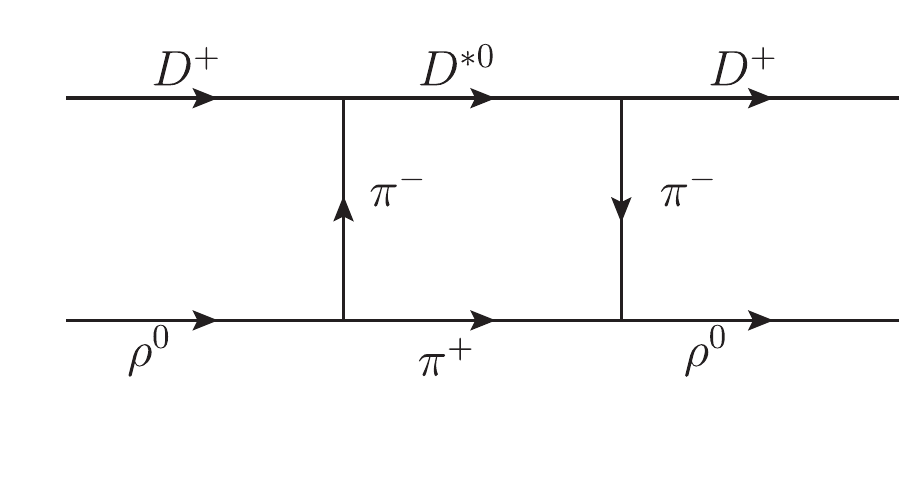}
    \caption{}
    \label{diag21}
\end{subfigure}
\quad
\begin{subfigure}{0.45\textwidth}
    \includegraphics[width=\textwidth]{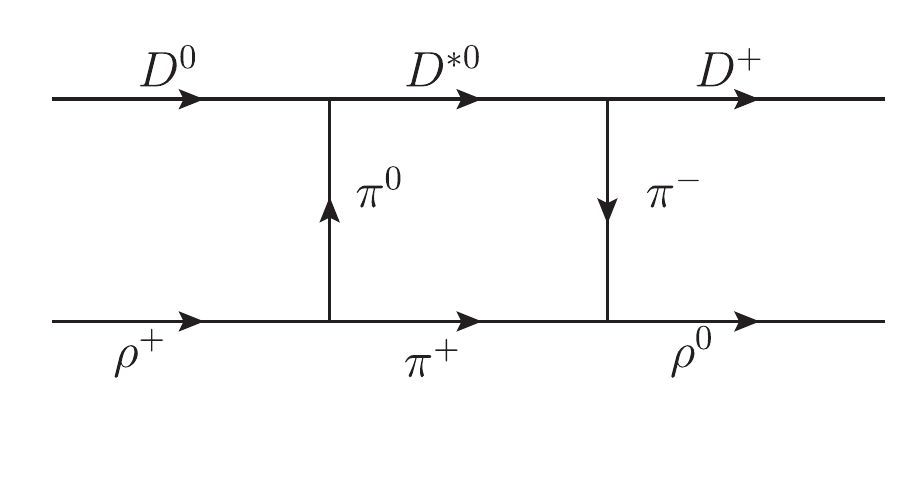}
    \caption{}
    \label{diag22}
\end{subfigure}
\caption{Box diagrams for the transitions $D^+\rho^0\to  D^{*0}\pi^+\to D^+\rho^0$ (left) and $D^0\rho^+\to  D^{*0}\pi^+\to D^+\rho^0$ (right).}
  \label{diag2}
\label{fig:figures}
\end{figure}

The box diagrams leading to non-zero contributions for  the mentioned transitions  are  shown in Figs.~\ref{diag1} and \ref{diag2} (with the momenta labels as  in Fig.~\ref{Drhocaixa}). 

We  use the following vector-pseudoscalar-pseudoscalar (VPP) Lagrangian
\begin{equation}
    \mathcal{L}_{VPP} = -ig_{VPP}\left<V_\mu\left[P,\partial^\mu P\right]\right>,\label{LVPP}
\end{equation}
to deduce the amplitudes for each of the diagrams presented in Figs.~\ref{diag1} and \ref{diag2}, 
where the coupling is related to the pion decay constant, $g_{VPP}=m_\rho/(2f_\pi)$, with $f_\pi=93$ MeV,
\begin{equation}
P =\left(\begin{array}{cccc}
        \frac{\pi^0}{\sqrt{2}} + \frac{\eta}{\sqrt{3}} + \frac{\eta'}{\sqrt{6}} & \pi^+ & K^+ & \Bar{D}^0 \\
        \pi^- & -\frac{\pi^0}{\sqrt{2}} + \frac{\eta}{\sqrt{3}} + \frac{\eta'}{\sqrt{6}}&K^0&D^-\\
        K^-&\Bar{K}^0&-\frac{\eta}{\sqrt{3}} + \sqrt{\frac{2}{3}}\eta'&D^-_s\\
        D^0&D^+&D^+_s&\eta_c
    \end{array}\right),
\end{equation}
and
\begin{equation}
    V_\mu=\left(\begin{array}{cccc}
    \frac{\rho^0}{\sqrt{2}}+\frac{\omega}{\sqrt{2}} & \rho^+&K^{*+}&\Bar{D}^{*0}\\
    \rho^-&-\frac{\rho^0}{\sqrt{2}}+\frac{\omega}{\sqrt{2}}&K^{*0}&D^{*-}\\
    K^{*-}&\Bar{K}^{*0}&\phi&D^{*-}_s\\
    D^{*0}&D^{*+}&D^{*+}_s&J/\psi
    \end{array}\right)_\mu.
\end{equation}

Consequently, we get the following common expression for the amplitudes of the diagrams  in Figs.~\ref{diag1} and \ref{diag2}
\begin{align}\nonumber
    -it^C_{D\rho\to D\rho} &= \xi^C\;\int \frac{d^4q}{(2\pi)^4}(\vec{\epsilon}_
    {D^*}\cdot\Vec{q}\,)^2 \vec{\epsilon}_{\rho}\cdot\Vec{q}~~\vec{\epsilon}^{\,\prime}_{\rho}\cdot\Vec{q}\, \left(\frac{1}{q^2-m^2_\pi + i\epsilon}\right)^2\\
    &\times\frac{1}{(P_2-q)^2-m^2_\pi + i\epsilon}~~\frac{1}{(P_1+q)^2-m^2_{D^*} + i\epsilon},\label{boxdiagramt}
\end{align}
where $\xi^C$ is a coefficient whose values are presented in Table \ref{coefcarga}. It is important to mention here that the expression in Eq.~(\ref{boxdiagramt}) is obtained  considering  $\vec{P}_1 = \vec{P}_2 \sim \vec{0}$, which is a fair approximation at energies close to the threshold of the reaction, as in our case. This latter consideration implies that  $P_1\sim P_1^\prime$, $P_2\sim P_2^\prime$, and consequently $q=q^\prime$.
\begin{table}[h!]
    \centering
    \caption{Coefficients $\xi^C$ present in the expression for the box diagrams in Eq.~(\ref{boxdiagramt}).}
    \begin{tabular}{c|c}\hline
    Transition&$\xi^C$\\\hline\hline
       $D^0\rho^+\to D^{*0}\pi^+\to D^0\rho^+$  & $4g^4$ \\
        $D^0\rho^+\to D^{*0}\rho^+\to D^+\rho^0$ & $-4\sqrt{2}g^4$\\
        $D^+\rho^0\to D^{*0}\pi^+\to D^+\rho^0$&$8g^4$\\
        $D^0\rho^+\to D^{*+}\pi^0\to D^0\rho^+$&$8g^4$\\\hline
    \end{tabular}
    \label{coefcarga}
\end{table}

Using  Eq.~(\ref{nabasedecarga}), along with the coefficients given in the Table~\ref{coefcarga}, we obtain the isospin projected amplitude as
\begin{align}
    -it^{box,\;I=1/2}_{D\rho\to D\rho} &=16\;g^4 \int\frac{d^4q}{(2\pi)^4}\;\Vec{q}\cdot\Vec{\epsilon}_{\rho}\;\Vec{q}\cdot\Vec{\epsilon'}_{\rho}\;\Vec{q}\cdot\Vec{\epsilon}_{D^*}\;\Vec{q}\cdot\Vec{\epsilon}_{D^*}\left(\frac{1}{q^2-m_{\pi}^2 + i\epsilon}\right)^2\nonumber\\
    &\times\left(\frac{1}{(P_1+q)^2-m_{D^*}^2 + i\epsilon}\right)\left(\frac{1}{(P_2-q)^2-m_{\pi}^2 + i\epsilon}\right).
\end{align}

The integration over the $q^0$ variable can be done analytically, by using Cauchy's residue theorem, to get
\begin{align}
    t^{box,\;I=1/2}_{D\rho\to D\rho} &= \frac{16}{3}g^4 \Vec{\epsilon}_\rho\cdot\Vec{\epsilon}^{\,\prime}_\rho\int\frac{d^3q}{(2\pi)^3}|\vec{q}\,|^4F^4(\vec{q})\left(\frac{M_{D^*}}{M_{K^*}}\right)^2\frac{1}{2\omega_{D^*}}\left(\frac{1}{2\omega_{\pi}}\right)^3\nonumber\\
    &\times\frac{1}{P_2^0+P_1^0-\omega_\pi - \omega_{D^*} + i\epsilon}\left[-\frac{1}{\omega_\pi}\left(\frac{1}{P_2^0 - 2\omega_\pi + i\epsilon} + \frac{1}{P_1^0 - \omega_\pi - \omega_{D^*} - i\epsilon}\right)\right.\nonumber\\
    &\left.+ \frac{1}{(P_2^0 - 2\omega_\pi - i\epsilon)^2} + \frac{1}{(P_1^0 - \omega_\pi - \omega_{D^*} + i\epsilon)^2}\right] + \mathcal B\label{eqtbox}
\end{align}
where $\omega_A = \sqrt{\vec{q}^{\,2} + M_A}$, with $A = \pi, D^*$ and 
\begin{align}
  \mathcal   B &= \frac{16}{3}g^4\Vec{\epsilon}_\rho\cdot\Vec{\epsilon}^{\,\prime}_\rho\int\frac{d^3q}{(2\pi)^3}|\vec{q}\,|^4F^4(\vec{q})\left(\frac{M_{D^*}}{M_{K^*}}\right)^2\frac{1}{2\omega_{D^*}}\left(\frac{1}{2\omega_{\pi}}\right)^3\nonumber\\
    &\times\left[\frac{1}{P_2^0 + 2\omega_\pi}\frac{1}{P_1^0 - \omega_\pi - \omega_{D^*}}\right]\left(\frac{1}{\omega_\pi} + \frac{1}{P_2^0 + 2\omega_\pi} - \frac{1}{P_1^0 - \omega_\pi - \omega_{D^*}}\right).
\end{align}
Notice that we have included a form factor in Eq.~(\ref{eqtbox}), which we take to be
\begin{align}
F(q) = \exp{\frac{(q^0)^2 - \vec{q}^{\,2}}{\Lambda^2}},
\end{align}
with $q^0=P^0_1-\omega_{D^*_s}$, at  each vertex and a factor $(M_{D^*}/M_{K^*})^2$ to account for the difference between the coupling constants $g$ for the $D^*\to D\pi$ and $K^ *\to K\pi$ vertices~\cite{Liang:2014eba}.   The value of the cut-off, $\Lambda$, is 1200 MeV. The remaining $d^3 q$ integral is done numerically, by setting the upper limit of $|\vec q\,|$ to 2000 MeV for practical purposes. We have verified that integrating to higher values of $|\vec q\,|$ does not change the results.  

It may be noticed that $\mathcal  B$ has no imaginary part because the expression $(P_2^0 + 2\omega_\pi)^{-1}$ is always positive and $(P_1^0 - \omega_\pi - \omega_{D^*})^{-1}$ is associated with   $D \to \pi + D^*$, a process which doesn't occur on-shell. Due to the mentioned fact, we don't have to worry about these terms having singularities and do not need to include an imaginary part ($i\epsilon$).

The resulting amplitude, $t^{box,\;I=1/2}_{D\rho\to D\rho}$, is added to the one related to a vector exchange of Ref.~\cite{gamermann2007}.  With this new amplitude we solve the Bethe-Salpeter equation, considering coupled channels as in Ref.~\cite{gamermann2007}. Apart from adding new diagrams, we have also adjusted the decay constant and the subtraction constant $\alpha$, which were set to $\sqrt{f_D f_\pi}$ and $-1.55$, respectively, in Ref.~\cite{gamermann2007}.

\begin{figure}[h!]
    \centering
    \includegraphics[width = 8cm]{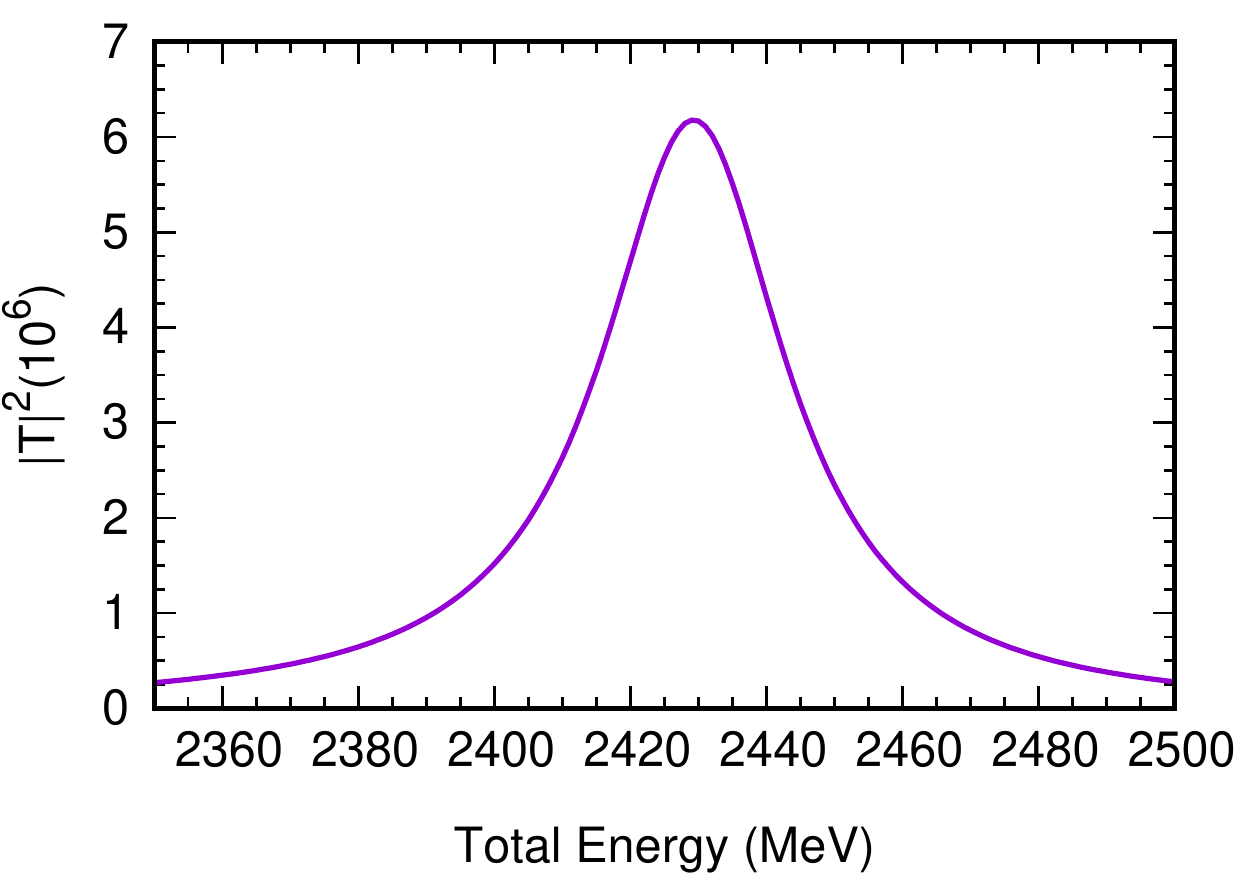}
    \caption{Modulus squared of the  $D\rho\to D\rho$ $t$-matrix, in isospin $1/2$, $J^P=1^+$. This result is obtained by considering the amplitude of Ref.~\cite{gamermann2007}  together with the contribution obtained from the box diagram shown in Fig. \ref{Drhocaixa} and solving  the Bethe-Salpeter equation in a coupled channel approach.}
    \label{modulotdoiscorpos}
\end{figure}

Using  $f_\pi$ as the decay constant, $\alpha = -1.45$, at the regularization scale $\mu = 1500$ MeV, (which is the same as in Ref.~\cite{gamermann2007}) we find that the modulus squared amplitude for $D\rho$ in isospin $1/2$, spin-parity $J^P=1^+$, peaks at  $\sim2428$ MeV, with a total width of $33$ MeV (see Fig.~\ref{modulotdoiscorpos}). 
Our findings are in excellent agreement with the mass and width values determined  from the most recent data obtained by the LHCb Collaboration~\cite{LHCb:2019juy},
 $M=2424.8\pm0.1\pm0.7$~MeV, $\Gamma =33.6\pm 0.3\pm 2.7$~MeV,  and by the BES Collaboration,  $M=2427.2\pm1.0\pm1.2$~MeV, $\Gamma =23.2\pm 2.3\pm 2.3$~MeV~\cite{BESIII:2019phe}.

We can, thus, summarize this discussion by mentioning that the addition of the amplitude obtained from the box diagram increases the width of the state found in Ref.~\cite{gamermann2007} by about $\sim 7$ MeV. Additionally, the model parameters have been fine tuned to better describe the properties of $D_1(2420)$, like the mass, which now is 2428 MeV instead of the value $\sim 2526$ MeV found in Ref.~\cite{gamermann2007}.

\section{The  $K\rho\bar{D}$ and  $K\rho D$ systems}

Inspired by the results obtained for the two-body system, $\rho D$, and recalling the fact that the $K\rho$, $KD$ and $K\bar{D}$  interactions are  attractive in nature, we find it encouraging to explore the systems $K \rho D$ and $K \rho \bar{D}$ (the results obtained for the $D\rho$ system must be equivalent to those for $\bar{D}\rho$, given the fact that $\rho$ is its own antiparticle and a system and its complex conjugate have equivalent descriptions).

\begin{figure}[h!]
    \centering
    \includegraphics[width=0.8\textwidth]{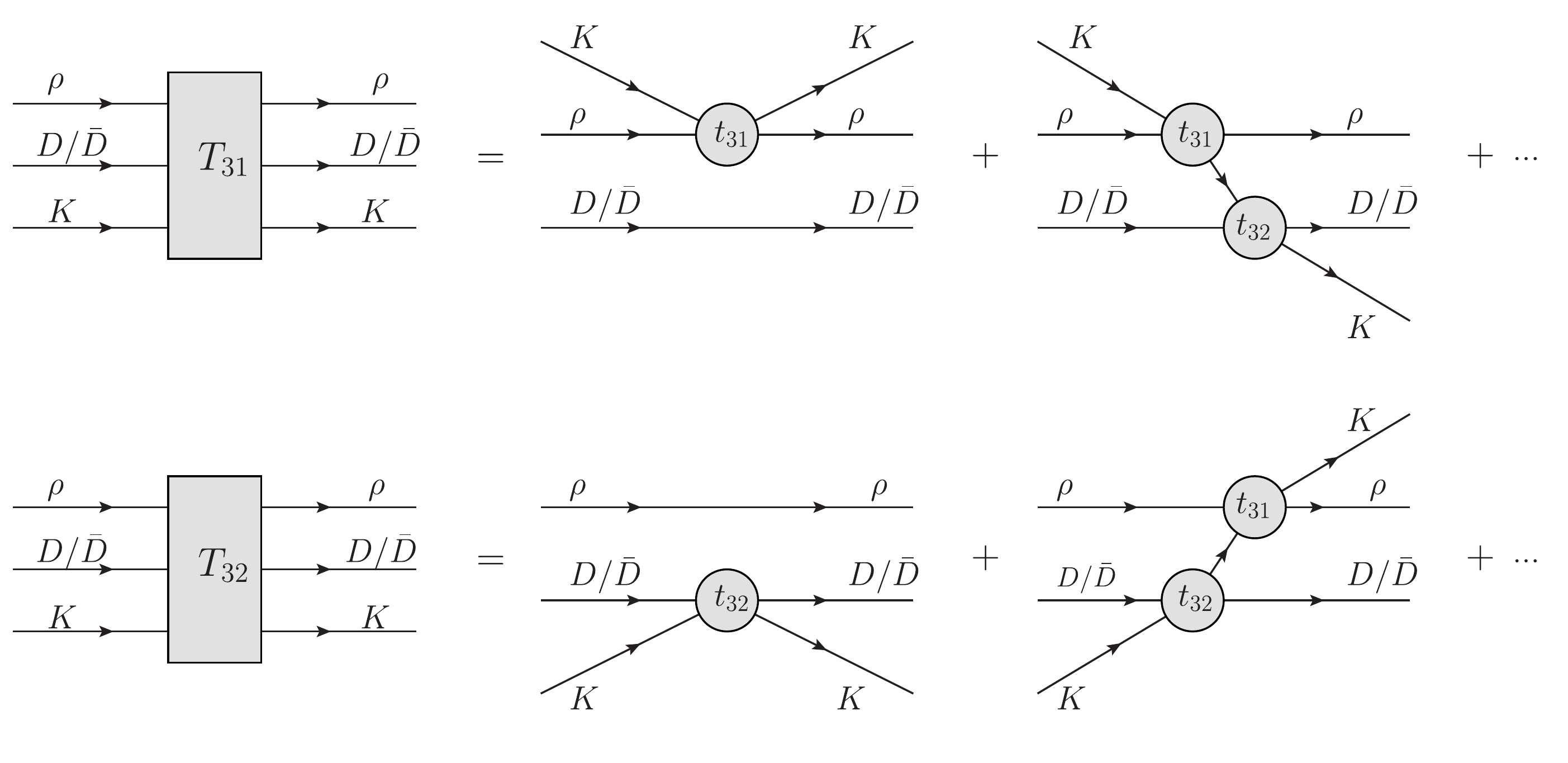}
    \caption{Diagrams for three-body interactions in the  $K\rho D$ $(K\rho \bar{D})$  system when treating $\rho \bar{D}$ $(\rho \bar{D})$ as a cluster.}
    \label{centrofixo}
\end{figure}

To describe the three-body interactions of the $K\rho D$ and $K\rho \Bar{D}$ systems we solve three-body equations within the static or fixed center approximation~\cite{Brueckner:1953zza,Barrett:1999cw,Kamalov:2000iy}. The fixed center approximation consists of considering that one of the particles (which is lighter than the other two) interacts with a cluster of the other two  strongly interacting particles, which remains unaltered in the scattering. We have a system at hand which is precisely suitable for the aforementioned treatment. The dynamics in the  $K\rho D \,(K\rho\bar{D})$ system can be described in terms of the interaction of the meson $K$ with the cluster formed by $\rho D \,(\rho \bar{D})$, which, as shown in the previous section, generates the state $D_1(2420)$. In other words, $D_1(2420)$ can be interpreted as a $\rho D$ quasi-bound state. The diagrams contributing to the scattering equations, within such a reorganization of the three-body system, can be drawn as shown in Fig.~\ref{centrofixo}, where the kaon rescatters off the constituents of the cluster.

The three-body  $T$-matrix is obtained by summing two coupled, infinite, scattering series
\begin{equation}
    T = T_{31} + T_{32},
\end{equation}
with
\begin{align}
    T_{31} = t_{31} + t_{31}G_{K}T_{32},\\
    T_{32} = t_{32} + t_{32}G_{K}T_{31},\label{T312}
\end{align}
where $t_{3i}$ represent the amplitude of the interaction of the particle labelled as 3 (kaon, in this case) with the $i$th particle in the cluster. The function $G_{K}$ represents the kaon propagating in the cluster and is given by the following expression~\cite{MartinezTorres:2020hus}
\begin{equation}
    G_{K} = 
    \int\frac{d^3q}{(2\pi)^3}\frac{F(\vec{q})}{(q^0)^2 - \vec{q}^{\,2} - m^2_K + i\epsilon},\label{propK}
\end{equation}
with $m_K$  being the mass of  kaon, and $q^0$ being the on-shell energy of the kaon in the rest frame of the cluster
\begin{equation}
    q^0 = \frac{s - m^2_K - M^2_c}{2 M_c},\label{q0}
\end{equation}
 where $M_c$ is the mass of the cluster. In  Eq.~(\ref{propK}), $F(\vec{q})$ is a form factor related to the wave function of the constituents of the cluster $\rho D\;(\rho \bar{D})$, and which is introduced to take into account the composite nature of the cluster
\begin{equation}
    F(\vec{q}) = \frac{1}{\mathcal{N}}\int_{|\vec{p}|,|\vec{p} - \vec{q}|<\Lambda}d^3\vec{p}f(\vec{p})f(\vec{p}-\vec{q}),
\end{equation}
where
\begin{equation}
    f(\vec{p}) = \frac{1}{\omega_{\rho}(\vec{p})\omega_{D(\bar{D})}(\vec{p})}\frac{1}{M_c-\omega_{\rho}(\vec{p}) - \omega_{D(\bar{D})}(\vec{p})},\label{funcestru}
\end{equation}
with $\mathcal{N}$ being a normalization factor defined such that $F(0)=1$, and  $\omega_A = \sqrt{m^2_A + \vec{p}^{\,2}}$, with $A = D(\bar{D}),\;\rho$.
The  cut-off $\Lambda$ is taken to be $\sim 960$ MeV, which is related to the value of the subtraction constant used to regularize the two-body loop functions in the $D \rho$ and coupled channels generating $D_1(2420)$. We shall vary this value in the range 900-1000 MeV to estimate the model uncertainties.

As shown in the previous section, $D_1(2420)$ [$\bar{D}_1(2420)$] is generated by the $\rho D\; (\rho \bar{D})$  and coupled channel interactions and has a mass $\sim 2428$ MeV. Hence, we use this value as the mass of the cluster in Eqs. (\ref{propK}) and (\ref{funcestru}). The  $D_1(2420)$ state also has a finite width, of approximately $33$~MeV, which is considered in the formalism by replacing $M_c\to M_c - i\frac{\Gamma_c}{2}$ in Eq.~(\ref{funcestru}).

To proceed with the calculations, we need to write the three-body system in a well defined isospin basis. Within the fixed-center approximation, the three-body system $K\rho D$ can be described as an effective two-body $K D_1(2420)$ system. Since both $D_1(2420)$ and $K$ have isospin $1/2$, the three-body system can have a total isospin $0$ or $1$. Considering the sign convention  defined in Eq.~(\ref{convsinal}), together with
\begin{equation}
    \ket{K} = \left(\begin{array}{c}
    \ket{K^+}\\
    \ket{K^0} 
    \end{array}\right),
\end{equation}
we can write the $K\rho D$/$K\rho\bar D$ systems with isospin $0$ or $1$ as:
\begin{align}\nonumber
    \ket{K\rho D, I = 0, I_3 = 0} &= -\frac{1}{\sqrt{6}}\ket{K^+\rho^0D^0} - \frac{1}{\sqrt{3}}\ket{K^+\rho^-D^+}
    -\frac{1}{\sqrt{3}}\ket{K^0\rho^+D^0} + \frac{1}{\sqrt{6}}\ket{K^0\rho^0D^+},\\\nonumber
   \ket{K\rho D, I=1, I_3=1}& = \sqrt{\frac{2}{3}}\ket{K^+\rho^+D^0} - \frac{1}{\sqrt{3}}\ket{K^+\rho^0D^+},\\\nonumber
    \ket{K\rho \bar{D}, I = 0, I_3 = 0} &= \frac{1}{\sqrt{6}}\ket{K^+\rho^0D^-} - \frac{1}{\sqrt{3}}\ket{K^+\rho^-\bar{D}^0}
    +\frac{1}{\sqrt{3}}\ket{K^0\rho^+D^-} + \frac{1}{\sqrt{6}}\ket{K^0\rho^0\bar{D}^0},\\
    \ket{K\rho \bar{D}, I=1, I_3=1} &= -\sqrt{\frac{2}{3}}\ket{K^+\rho^+D^-} - \frac{1}{\sqrt{3}}\ket{K^+\rho^0\bar{D}^0}.\label{KrhobarD11}
\end{align}
However,  to calculate the three-body $T$-matrix within the FCA we must determine the  $t_{31}$ and $t_{32}$ amplitudes, which describe, respectively, the interaction between a particle, labelled as $3$ (in this case the kaon), with the particles, labelled as 1 ($\rho$) and 2 ($D$, or $\bar D$), of the cluster, respectively. It is then more convenient to use states written in terms of the isospin of the (31) or (32) subsystems. To better explain the isospin dependence, we consider the specific example of total isospin $I=0$ of the three-body system. For this purpose, we use  Clebsch-Gordan coefficients to rewrite the first equation in the set labelled as Eq.~(\ref{KrhobarD11}) in terms of the isospin of the $K\rho$ system. We can write, for $\ket{K^+\rho^0}$,
\begin{align}
\ket{K^+\rho^0}&=\ket{K,I=\frac{1}{2},I_3=\frac{1}{2}}\otimes\ket{\rho, I=1,I_3=0}\nonumber\\
&=\sqrt{\frac{2}{3}}\ket{K\rho, I=\frac{3}{2},I_3=\frac{1}{2}}+\frac{1}{\sqrt{3}}\ket{K\rho, I=\frac{1}{2},I_3=\frac{1}{2}},\label{Krhoket}
\end{align}
and similar equations for $\ket{K^+\rho^-}$, $\ket{K^0\rho^+}$, $\ket{K^0\rho^0}$.  Substituting such expressions in Eq.~(\ref{KrhobarD11}), we find 
\begin{align}
    \ket{K\rho D, I = 0, I_3=0} &= \frac{1}{\sqrt{2}}\Big[\ket{K\rho,I=1/2,I_3=1/2}\otimes\ket{D,I=1/2,I_3=-1/2}\nonumber\\
    &- \ket{K\rho,I=1/2,I_3=-1/2}\otimes\ket{D,I=1/2,I_3=1/2}\Big],\label{rhoKbase}
\end{align}
which can be used to determine $t_{31}$,
\begin{align}
    t_{31} = \bra{K\rho D,I=0,I_3=0}t\ket{K\rho D,I=0,I_3=0} =t^{1/2}_{K\rho},
\end{align}
where $t^{1/2}_{K\rho}$ is the two-body $t$-matrix for the $K \rho$ system in isospin $1/2$. For other total isospin values, $t_{31}$ is a combination of the $K\rho$ two-body $t$-matrices in isospin $1/2$ and $3/2$, with weights determined from products of Clebsch-Gordan coefficients. Similarly, writing $\ket{K\rho D,I=0,I_3=0}$ in terms of the isospin of the $KD$ system, we can evaluate $t_{32}$, which, in general, can be written in terms of combinations of the $KD$ two-body $t$-matrices in isospin 1 and 0. To express this information, we introduce a compact notation by writing 
\begin{align}
    t_{31} &= \bra{K\rho \mathds{D}, I, I_3 }t\ket{K\rho \mathds{D},I,I_3}\equiv \vec{\omega}_{31}^{I}\cdot\vec{t}_{31}\label{t31II3}\\
    t_{32} &=\bra{K\rho \mathds{D},I,I_3}t\ket{K\rho \mathds{D},I,I_3}\equiv \vec{\omega}_{32}^{I}\cdot\vec{t}_{32},\label{t32II3}
\end{align}
where $\mathds{D} = D,\;\bar{D}$ when considering the system $K\rho D$ or $K\rho \bar{D}$, respectively, and $I$ is the total isospin of the three-body system. The vectors $\omega^{I}_{31\;(32)}$ are the weight factors related to Clebsch-Gordan coefficients, which are summarized in Table~\ref{combisospin},  and $\vec{t}_{31}$ and $\vec{t}_{32}$ are defined as follows
\begin{equation}
    \vec{t}_{31} \equiv \left(\begin{array}{c}
    t^{1/2}_{K\rho}\\
    t^{3/2}_{K\rho}
    \end{array}\right)\phantom{nana}\vec{t}_{32} \equiv \left(\begin{array}{c}
    t^{0}_{K(D/\bar{D})}\\
    t^{1}_{K(D/\bar{D})} 
    \end{array}\right).
\end{equation}

\begin{table}[h!]
    \centering
    \caption{$\omega_{31}$ and $\omega_{32}$ for the $K\rho D\; (K\rho \bar{D})$ system for total isospin $0$ and 1.}
    \begin{tabular}{c|c|c}\hline
    Total isospin ($I$) &$\vec{\omega}^{I}_{31}$&$\vec{\omega}^{I}_{32}$\\\hline\hline
    0&$(1,\;0)$&$(0,\;1)$\\
    1&$({1}/{9},\;{8}/{9})$& $({1}/{3},{2}/{3})$\\\hline
    \end{tabular}
    \label{combisospin}
\end{table}

It can be noticed that to determine $t_{31}$ and $t_{32}$ we need the two-body $t$-matrices of the $K\rho$, $KD/K\bar D$ systems in the different isospin configurations. These latter amplitudes are obtained by solving Bethe-Salpeter equations, considering all coupled channels relevant in each case, keeping all interactions in $s$-wave.

To obtain the  two-body $t$-matrices for the $K\rho$ and coupled channels we follow Refs.~\cite{Roca:2005nm,Geng:2006yb}, where the formalism is built by considering  vector mesons as fields transforming homogeneously under the nonlinear realization of chiral symmetry. The model shows that the amplitudes lead to the formation of $K_1(1270)$, which is related to two poles in the complex energy plane, and well describes the data on the $K^- p\to K^- \pi^+\pi^- p$ process.  

For the  $KD$ ($K\bar{D}$ ) subsystems and coupled channels we consider as input for the Bethe-Salpeter equation an amplitude obtained from a Lagrangian based on the heavy quark spin symmetry~\cite{Guo:2006fu}. In this case, the interactions in the charm and strangeness $+1$ isoscalar system generate $D_s(2317)$ while the isoscalar $K\bar{D}$ interaction is found to be weakly attractive in nature. It should be mentioned that the strong relation between the $KD$ system and $D_s(2317)$ has been confirmed by  analysis of the lattice data too~\cite{MartinezTorres:2014kpc,MartinezTorres:2017bdo,MartinezTorres:2011pr}.

As we have already mentioned, the interactions of the three particles that compose the systems $K\rho D$ and $K\rho \bar{D}$ can be interpreted as the interaction of the kaon with a cluster. If we compare the expression for the  three-body $S$-matrix with that for an effective, two-body, kaon-cluster scattering, we find that in order for  both $S$-matrices to be compatible we must redefine $G_{K}$, $t_{31}$ and $t_{32}$, in Eq.~(\ref{T312}), as  follows~\cite{MartinezTorres:2010ax,Roca:2010tf,Ren:2018pcd}
\begin{align}
    G_{K}\to \frac{1}{2M_c}G_{K},\phantom{na}t_{31} \to \frac{M_c}{m_{\rho}}t_{31},\phantom{na}
    t_{32} \to \frac{M_c}{m_{D(\bar{D})}}t_{32}.
\end{align}

Having described all inputs necessary for the determination of  the three-body $T$-matrices for the systems $K\rho D$ and $K\rho \bar{D}$ in isospin $1$ and $0$, we are in a position to discuss the results. Let us begin by discussing the results for the $K\rho \bar{D}$ system, which has the quantum numbers of $X_1$ ($J^P = 1^-$, $C=-1$, $S=+1$) found by the LHCb Collaboration~\cite{LHCb:2020bls,LHCb:2020pxc}.
\begin{figure}[h!]
    \centering
    \includegraphics[width=9cm]{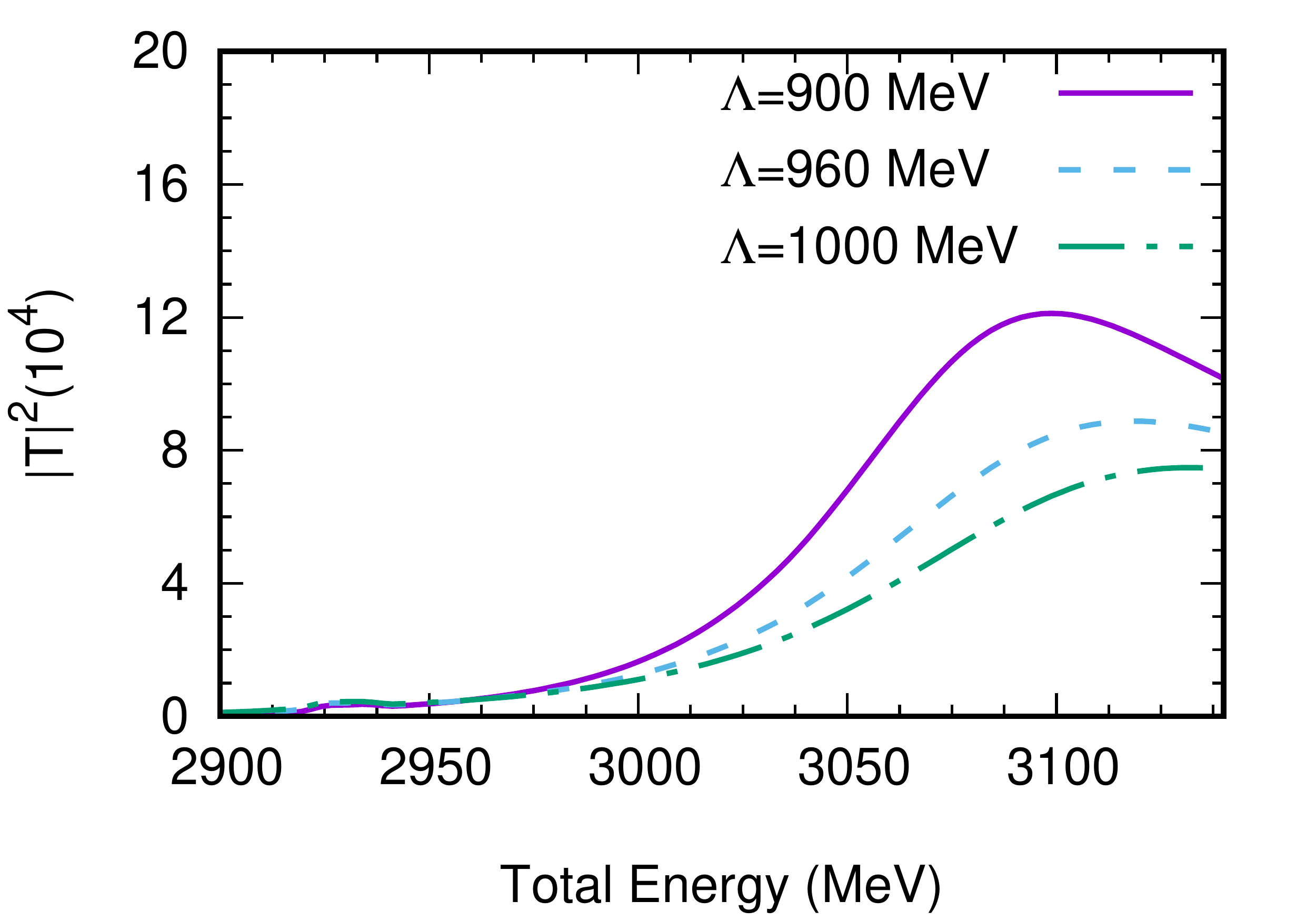}
    \caption{Modulus squared of the $T$-matrix for the  $K\rho \bar{D}$ system with total isospin $0$.}
    \label{kbarDrho0}
\end{figure}
 The results are shown in Fig.~\ref{kbarDrho0} for total isospin zero, for three different values of the cut-off used to calculate Eq.~(\ref{propK}). 
 
The results are depicted up to the energy corresponding to the three-body threshold for twofold reasons. One is that we are looking for a  $K\rho \bar D$ bound state, which can be associated with the $X_1$ state.  Yet other reason is that $\bar D$ and $\rho$, recalling the their interaction leads to the formation of $D_1(2420)$, as shown in section~\ref{sec:2}, are considered as static or  fixed scattering centers. It was shown in Ref.~\cite{MartinezTorres:2010ax} that the results in such a formalism are reliable at energies below the three-body threshold. Discussions on the applicability of the formalism can also be found in Refs.~\cite{Xie:2010ig,Malabarba:2021taj}. Coming back to the results shown in Fig.~\ref{kbarDrho0}, it can be said that a bump seems to appear for the value of cut-off 1000 MeV, with the mass $\sim 3085$ MeV.  Such a mass value, however, does not agree with the one determined by LHCb for the $X_1$ state ($2904 \pm 5$ MeV).

\begin{figure}[h!]
    \centering
    \includegraphics[width=9cm]{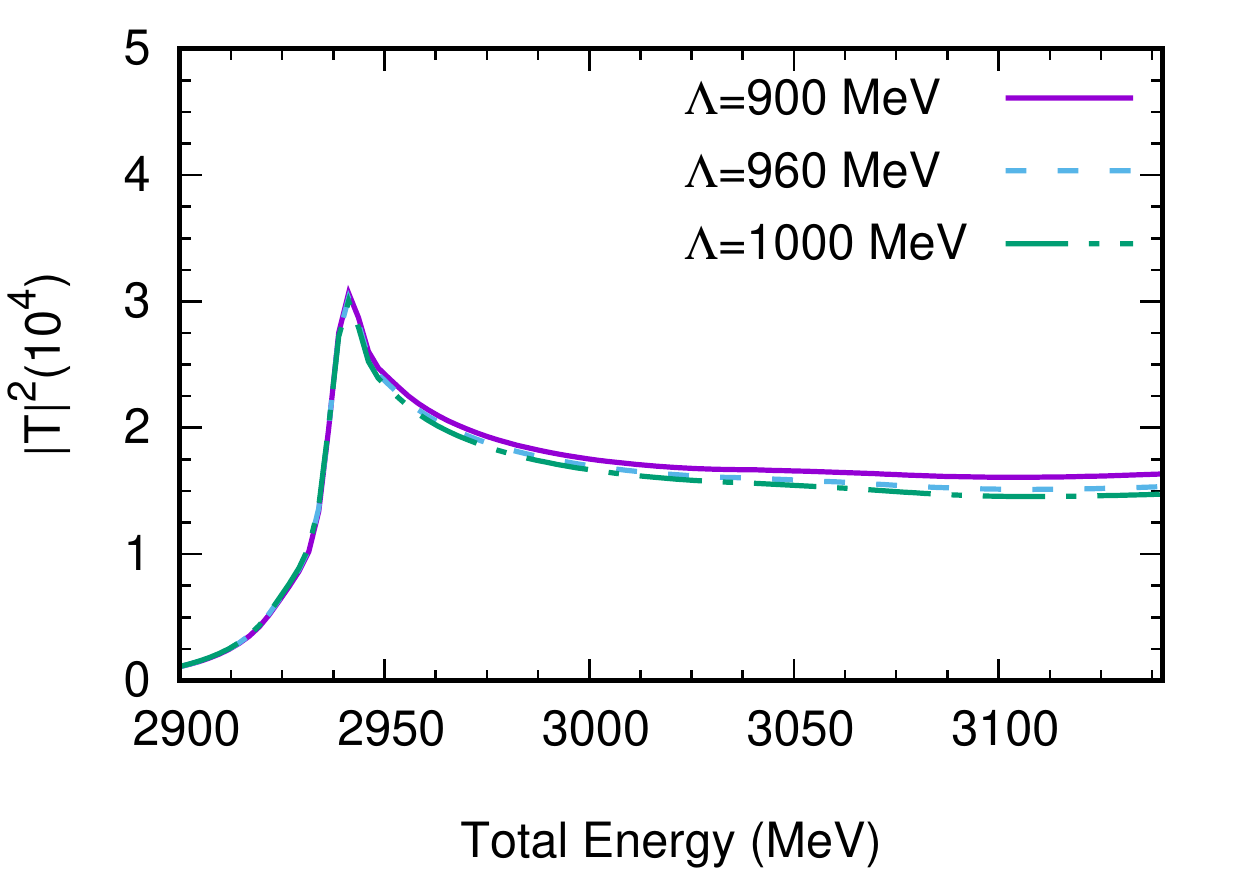}
    \caption{Modulus squared of the three-body $T$-matrix for the  $K\rho \bar{D}$ system with total isospin $1$.}
    \label{kbarDrho1}
\end{figure}

Let us now look at Fig.~\ref{kbarDrho1}, which shows  the $K\rho \bar{D}$ amplitude in total isospin 1.
Clearly, such a configuration does not form a resonance and only a cusp is seen around the opening of the $K\bar D_1(2420)$ threshold. We can, thus, conclude that neither of the isospin configurations of $K\bar D_1(2420)$ forms a state which can be related to the $X_1$ state of LHCb~\cite{LHCb:2020bls,LHCb:2020pxc}. 

Let us now compare our results with  those found in Refs.~\cite{Dong:2020rgs,He:2020btl,Qi:2021iyv,Chen:2021tad}. It is reported in Ref.~\cite{Dong:2020rgs} that the  $K \bar D_1(2420)$ system does not bind, neither in the isoscalar nor in the isovector configuration. Our results agree with  those in Ref.~\cite{Dong:2020rgs}.  We  find only a bump appearing in the isoscalar $K \bar D_1(2420)$ system above its threshold but below the three-body ($K\rho \bar D$) threshold. Recall that  Refs.~\cite{Dong:2020rgs,He:2020btl,Qi:2021iyv,Chen:2021tad} treat $\bar D_1(2420)$  as an effective field and study a two body $K \bar D_1(2420)$ system. We, on the other hand, treat $\bar D_1(2420)$ as a $\bar D \rho$ molecular state and study  a three-body system, $K \rho  \bar D$.  Our findings also coincide with those in Ref.~\cite{He:2020btl}, where only the isoscalar configuration is found to be attractive and requires a very large cut-off to form a bound state. As already stated before, we find a bump like structure in the isoscalar $K \bar D_1(2420)$ system above its threshold, and not below.
Our results, however, do not agree with those in Refs.~\cite{Qi:2021iyv,Chen:2021tad}. In   Ref.~\cite{Qi:2021iyv}, the $K \bar D_1(2420)$ system is found to bind in both isospin configurations, with the binding energy varying in the range $-5$ MeV to $-30$ MeV when changing the form factors and cut-off values. The width of the isovector state is found to be narrower ($\Gamma \sim$~12-30 MeV) when compared with the isoscalar state ($\Gamma\sim$ 60-100 MeV). Thus, it is concluded in Ref.~\cite{Qi:2021iyv} that their isoscalar state can be related with $X_1$. It is concluded in Ref.~\cite{Chen:2021tad} too that  $X_1$ can be interpreted as a $K\bar D_1(2420)$ bound state. As discussed above, we do not find formation of a state below the $K \bar D_1(2420)$, and cannot find a description for the $X_1$ of Refs.~\cite{LHCb:2020bls,LHCb:2020pxc}.

\begin{figure}[h!]
    \centering
    \includegraphics[width=8cm]{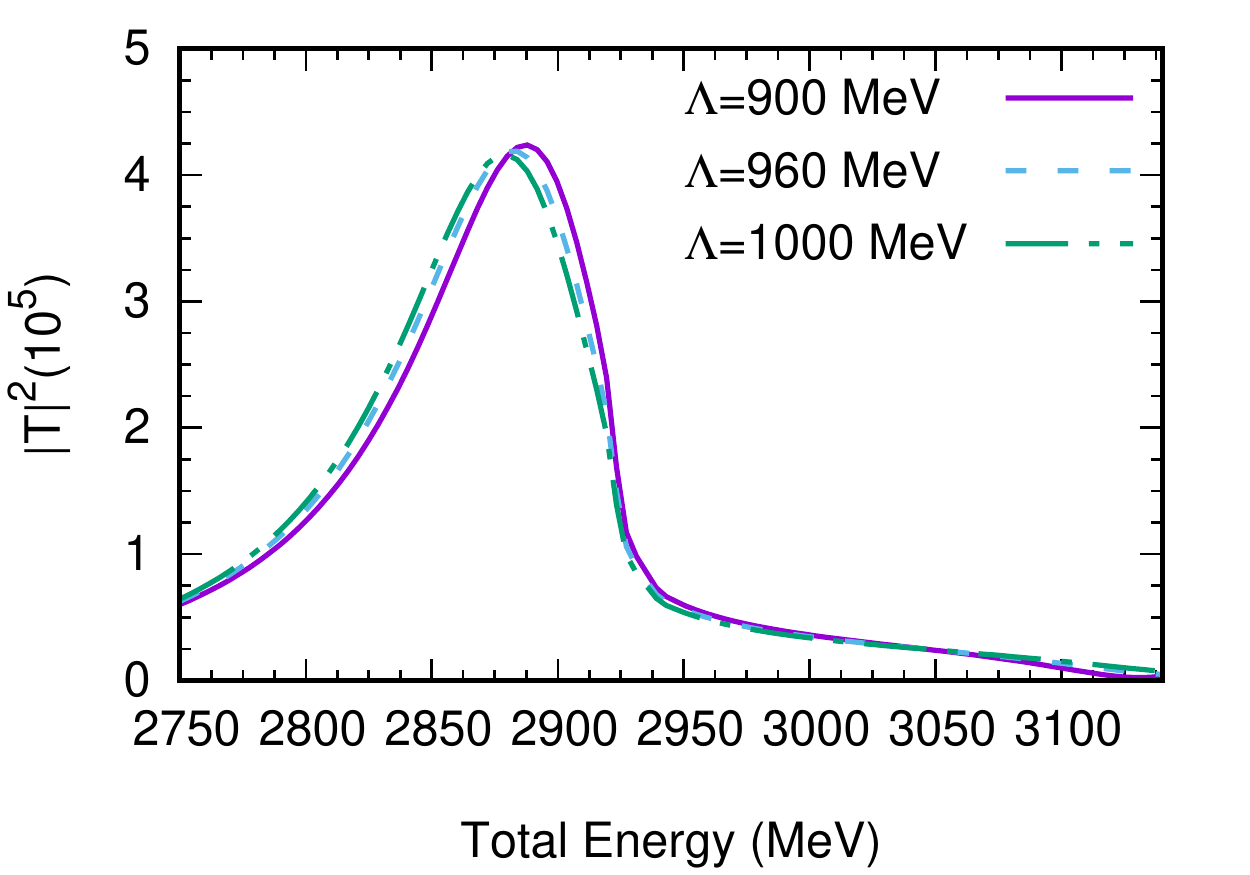}
    \caption{Modulus squared of the  $T$-matrix for the  $K\rho D$ system with total isospin $0$.}
    \label{kDrho0}
\end{figure}

Next we discuss the results obtained  for the $K\rho D$ system. The difference in this case is that all the subsystems interact strongly and lead to formation of a molecular state in one of the two-body isospin configurations. The $KD$ system forms an isoscalar bound state, $D_s(2317)$; $K\rho$ and coupled channels generate $K_1(1270)$, and as discussed in the previous section, $D_1(2420)$ can be generated  from  $\rho D$ and coupled channel interactions.  We show the modulus squared amplitude for the  $K\rho D$ system in total isospin 0 in Fig.~\ref{kDrho0}.
A clear peak can be seen with a mass around  $2872$ MeV and a width of about $100$~MeV. The quantum numbers associated with this state are $I(J^P) = 0\,(1^-)$ and it has a mass and width in excellent agreement with the known state $D_{s1}^*(2860)$~\cite{pdg}, with   $M=2859\pm 27$ MeV and $\Gamma =159\pm 80$ MeV determined by  LHCb~\cite{LHCb:2014ott}.  It is important to mention here that the state seen in Fig.~\ref{kDrho0} has a finite width, even though its  mass lies below the three-body threshold. This is so because lighter channels like  $K \pi D_s$ are implicitly present in the system. The presence of such open channels arises through the imaginary part of the two-body      $t$-matrices, which have been calculated in a coupled channel approach. To compare our results with other works, we recall that  the $K_1(1270)D$ and $K D_1(2420)$ systems were studied separately (uncoupled to each other) in Ref.~\cite{Dong:2020rgs}. The formalism in this former work was built by writing  an effective field for the axial mesons and the $K_1(1270)D$ system was found to form a bound state with mass 3112 MeV, while $K D_1(2420)$ was found to be weakly attractive.  In our case, the three-body system acts simultaneously as  $K_1(1270)D$ and $K D_1(2420)$ effective systems. Both the systems have similar mass and could be treated as coupled channels in the formalism of Ref.~\cite{Dong:2020rgs}. It sounds plausible that  such a treatment  could  lower the mass of  the state found in the  $K_1(1270)D$ system in Ref.~\cite{Dong:2020rgs} since the introduction of a coupled channel usually leads to a more attractive interaction (see section 6 of Ref.~\cite{Aceti:2014ala}).

Finally, we show the modulus squared amplitude for the $K\rho D$ system with total isospin $1$.  
\begin{figure}[h!]
    \centering
    \includegraphics[width=10cm]{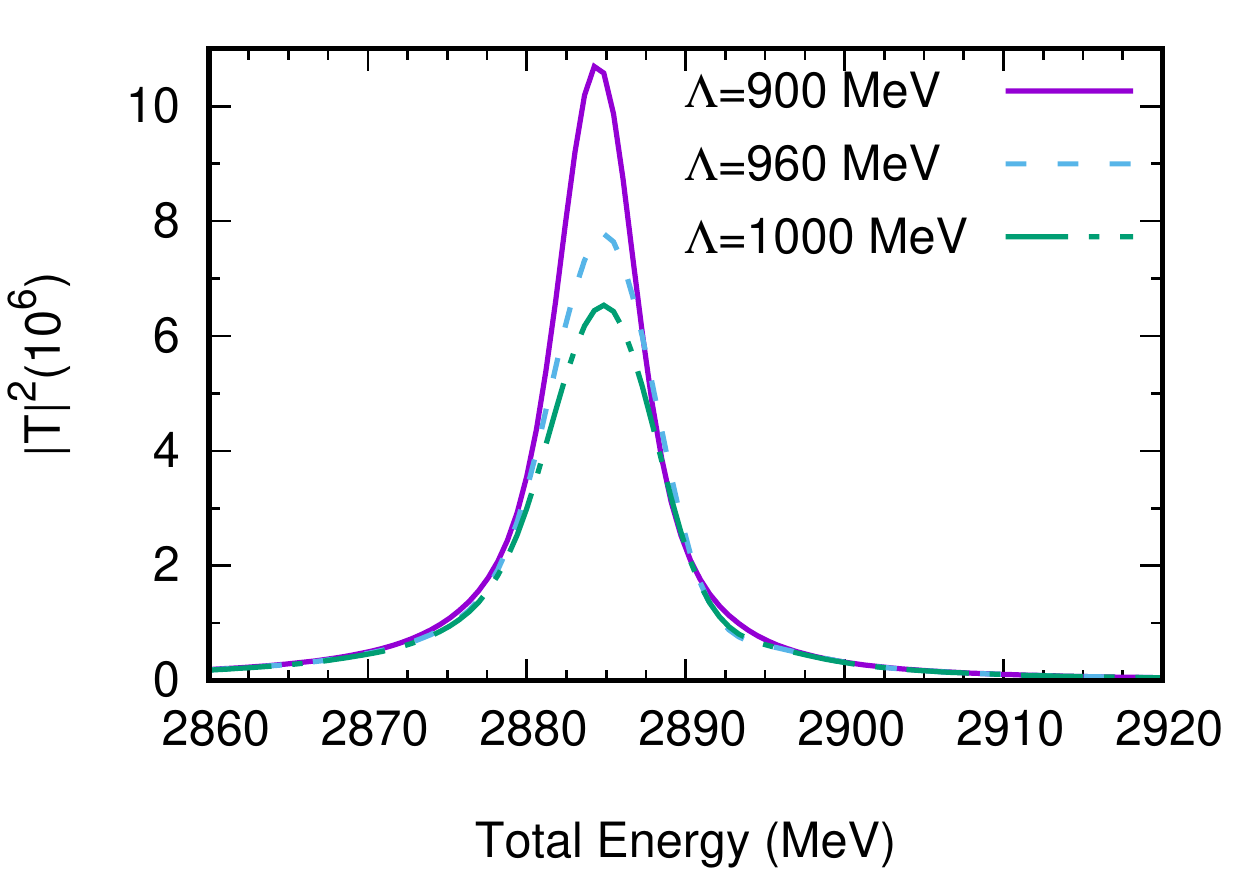}
    \caption{Modulus squared of the three-body $T$-matrix for the  $K\rho D$ system with total isospin $1$.}
    \label{kDrho1}
\end{figure}
A narrow and well pronounced peak is seen in Fig.~\ref{kDrho1}, which is very interesting since the corresponding state must have quantum numbers $I(J^P) = 1\,(1^-)$ with charm and strangeness $+1$. Such a state cannot be described as a $q\bar{q}$ state and the higher charged components of the multiplet necessarily require quarks with at least four different flavors in its wave function. The state is, thus, explicitly exotic and would be $c s\bar q\bar q$-like partner  of the  $\bar c \bar s q q$-like $X_i$ states found by LHCb~\cite{LHCb:2020bls,LHCb:2020pxc}. The state in Fig.~\ref{kDrho1} has a mass of  $\sim 2883$ MeV and width of $\sim 7.4 \pm 0.9$~MeV. It can be noticed that though the mass of our $K\rho D$ isovector state is similar to $X_i$s, its width is a lot narrower. In fact it might look surprising that the same system in isospin 0 produces a much wider state. The reason for such a difference is that the isoscalar $KD$ interaction does not contribute to the three-body interactions with total isospin 0, but does  contribute (generating $D_s(2317)$) to the total isospin 1 (see  Table~\ref{combisospin}). In other words, $K\rho D$ with total isospin 1  reorganizes itself, partly, as an effective $\rho D_s(2317)$ system and the mass of the three-body bound state lies about 200 MeV below the  $\rho D_s(2317)$ threshold. The same system gets contribution from  open channels like $K\pi D_s$ through the imaginary part of the isovector $DK$ interactions, which produces a finite, though small, width. Such a state has not been found yet and we hope that our findings will encourage its study in future experiments.

\section{Further investigations: $\pi X_i$ interactions}
Considering that $X_0$ is proposed to be a $\bar K^* D^*$ molecular state in most works~\cite{Molina:2020hde,Liu:2020nil,Chen:2020aos,Huang:2020ptc,Hu:2020mxp,Agaev:2020nrc,Chen:2021erj,Kong:2021ohg}, we find it useful to investigate if a pion with $X_0$ can form a bound state. Particularly, we follow Ref.~\cite{Molina:2020hde} where the $\bar K^* D^*$  system has been found to form isoscalar bound states with $J^P =0^+$, $1^+$ and $2^+$. Our main aim is to test the possibility of the existence of a bound state of pion and the $1^+$ state of the latter work, since the masses of  $X_i$s discovered by LHCb differ by less than the mass of a pion.

We, thus,  study the three-body system $\pi \bar{K}^* D^*$, using the same formalism as presented in the previous section  for $K\rho D$ and $K\rho\bar{D}$.  In this case we consider $\bar{K}^*D^*$ to form a cluster and that the pion scatters off its components. Let us denote the three states generated by the $\bar{K}^*D^*$ interactions in Ref.~\cite{Molina:2020hde} as $\tilde X_0$, $\tilde X_1$ and $\tilde X_2$, where $\tilde X_0$ has $I(J^P) = 0\,(0^+)$ with a mass of $2866$ MeV and width $\sim 57$ MeV and which is well identified with the $X_0$ state of LHCb. The other states, $\tilde X_1$ has $I(J^P) = 0\,(1^+)$, a mass of $2861$ MeV and about $20$ MeV of width, and $\tilde X_2$ has $I(J^P)=0\,(2^+)$, a mass of $2775$ MeV and a width of $38$ MeV. We study the three possible configurations of $\pi \bar{K}^* D^*$: $\pi\tilde X_0$, $\pi\tilde X_1$ and  $\pi\tilde X_2$. All the three system have total isospin 1, since the  $\bar{K}^* D^*$ subsystem clusters with total isospin $0$, and have the spin parity $0^-$, $1^-$ and $2^-$, respectively.

Considering the findings of Ref.~\cite{Molina:2020hde} to describe the cluster and determining the amplitudes for the  $\pi \bar{K}^*$, $\pi D^*$  subsystems by solving the Bethe-Salpeter equation following Refs.~\cite{Geng:2006yb,gamermann2007}, we determine the three-body $T$-matrix. We must emphasize that the amplitude for $\pi D^*$ is obtained by considering $D\rho$ and other coupled channels, by including the box diagrams, as described in section~\ref{sec:2}. The coupled channel system generates $D_1(2420)$ though, as mentioned in section~\ref{sec:2}, it couples weakly to $\pi D^*$. Similarly, $\pi \bar{K}^*$ is a coupled channel of $K\rho$ and other pseudoscalar-vector meson systems which generate $K_1(1270)$.

\begin{figure}[h!]
    \centering
    \includegraphics[width=10cm]{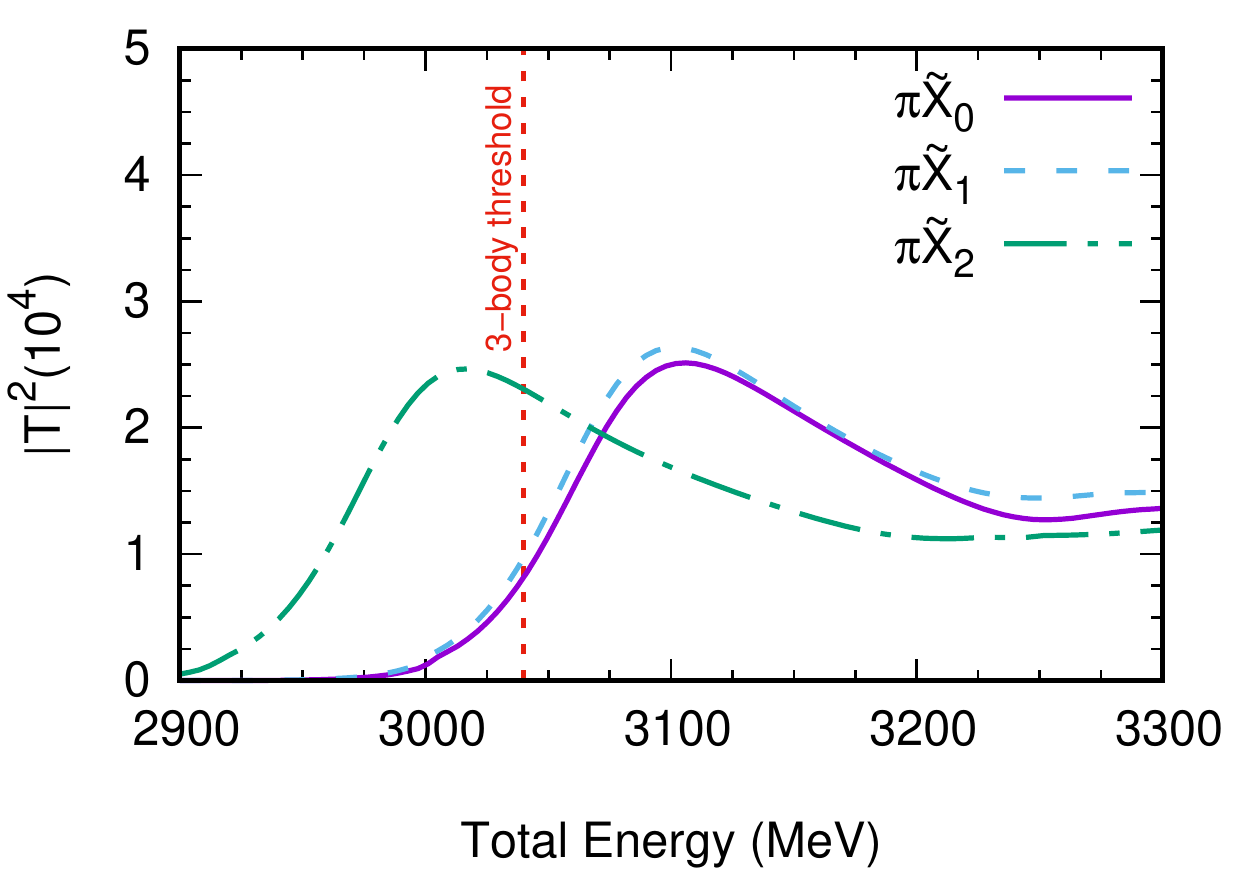}
    \caption{Modulus squared of the three-body $T$-matrix for the  $\pi \bar{K}^*D^*$ system considering $\bar{K}^*D^*$ to cluster as  $\tilde X_0$, $\tilde X_1$ and $\tilde X_2$ of Ref.~\cite{Molina:2020hde}.}
    \label{piontmatriz}
\end{figure}

We present the modulus squared three-body amplitudes in Fig.~\ref{piontmatriz} for the three possible total spins of the system. As can be seen, a three-body bound state with mass around 2900 MeV is not found in the  $\pi \bar{K}^* D^*$ system with isospin $1$. 
The $\pi \tilde X_2$ amplitude is seen to increase with the energy, indicating that a state may get formed at higher energies. We have extended our calculations to higher energies and find that a very wide bump like structure appears in the squared amplitude, at energies above the three-body threshold. We would, however, not make any claims since the calculations above the three-body threshold may not be reliable with our present formalism.

We can conclude this section by mentioning that the $X_1$ of LHCb cannot be described as $\pi \bar K^* D^*$ bound state.

\section{Summary and future perspectives}
In the present work we have investigated the possibility of describing the $X_1(2900)$ found by LHCb~\cite{LHCb:2020bls,LHCb:2020pxc} as a molecular state of $K \rho \bar D$ or $\pi \bar{K}^*D^*$. We first argue that the $D\rho$ interaction is strongly related to $D_1(2420)$ as first proposed in Ref.~\cite{gamermann2007}. We update this former work by including box diagrams with a pseudoscalar exchange  and by adjusting the model parameters to obtain the properties of  $D_1(2420)$ in better agreement with the latest data. We then study three-body systems by treating $D\rho$ ($\bar D\rho$) as components of $D_1(2420)$, which remain static in the scattering. Within such an approximation, we find that the $K \rho \bar D$
system could form a wide state but with a mass higher than that of the $X_1$ state found by  LHCb~\cite{LHCb:2020bls,LHCb:2020pxc}. We, thus, conclude that $X_1$ must have a description different than a $K \rho \bar D$ molecule.

Taking the advantage of the symmetry between $\bar D\rho$ and $D\rho$ interactions, we study the $K \rho D$ system. In this case, we find that the system generates states in the total isospin 0 as well as 1 configurations. The properties of the  isoscalar state turn out to be in excellent agreement with those of $D_{s1}^*(2860)$~\cite{pdg}, implying that it can be understood as a three-body molecular resonance. The isovector state is clearly exotic, requiring more than two quarks for its description. This latter state is narrow, having a width less than 10 MeV, and a mass similar to that of the $X_i$s found by LHCb.

 In the case of $\pi \bar{K}^*D^*$, we follow Ref.~\cite{Molina:2020hde} where $\bar{K}^*D^*$ interactions have been studied thoroughly, providing a molecular description for $X_0(2900)$~\cite{LHCb:2020bls,LHCb:2020pxc}, and proposing the existence of two other states with spin parity $1^+$, $2^+$. The investigations have been further  extended \text{in} Refs.~\cite{Dai:2022qwh,Dai:2022htx} suggesting reactions to observe the new predicted states. Treating $\bar{K}^*D^*$ as states with three possible spin-parities, as found in Ref.~\cite{Molina:2020hde}, we study  
 the $\pi \bar{K}^*D^*$ interaction. We do not find a state which can be associated with $X_1$, although bump like structures are found at energies around $3000-3100$ MeV. Such bumps lie in the energy region beyond the applicability of the FCA and a more detailed study would be necessary to make more robust claims.
  
As future perspectives of the current study, it should be useful to  investigate further the properties of the states found in this work. For example, it can be worthwhile to determine the decay rates of the $K\rho D$ and $K \rho \bar D$ states, with isospin 0, to different possible final states. The former state is associated with $D_{s1}^*(2860)$ in our work, for which not much is known on its decay properties. A theoretical study of such properties can be helpful in experimental investigations of the properties of $D_{s1}^*(2860)$.  
 
\section*{Acknowledgement}
B.B. M.,  K.P.K and A.M.T gratefully acknowledge the  support from the Funda\c c\~ao de Amparo \`a Pesquisa do Estado de S\~ao Paulo (FAPESP), processos n${}^\circ$ 2020/00676-8,   2019/17149-3 and 2019/16924-3. K.P.K and A.M.T are also thankful to the Conselho Nacional de Desenvolvimento Cient\'ifico e Tecnol\'ogico (CNPq) for grants n${}^\circ$ 305526/2019-7 and 303945/2019-2. This work is partly supported by the Spanish Ministerio de Econom\'ia y Competitividad and European
FEDER funds under Contracts No. PID2020-112777GBI00, and by Generalitat Valenciana under contract PROMETEO/2020/023. This project has received funding from the European Unions 10 Horizon 2020 research and innovation programme under grant agreement No. 824093 for the ``STRONG-2020" project.

\end{document}